\documentclass[a4paper]{jpconf}
\usepackage[dvipdfm]{graphicx}
\usepackage{wrapfig}
\usepackage{amsmath,amssymb}

\newcommand{\bra}[1]{\langle {#1} |}
\newcommand{\ket}[1]{| {#1} \rangle}

\begin{document}
\title{Real-time calculations of many-body dynamics in quantum systems
}

\author{Takashi Nakatsukasa}

\address{RIKEN Nishina Center, Wako, 351-0198, Japan}

\ead{nakatsukasa@riken.jp}

\begin{abstract}
Real-time computation of time-dependent quantum mechanical
problems are presented for nuclear many-body problems.
Quantum tunneling in nuclear fusion at low energy
 is described using a time-dependent wave packet.
A real-time method of calculating strength functions
using the time-dependent Schr\"odinger equation is utilized
to properly treat the continuum boundary condition.
To go beyond the few-body models,
we resort to the density-functional theory.
The nuclear mean-field models are briefly reviewed
to illustrate its foundation and necessity of state dependence in
effective interactions.
This state dependence is successfully taken into account by the density
dependence, leading to the energy density functional.
Photoabsorption cross sections in $^{238}$U are calculated with
the real-time method for the time-dependent density-functional theory.
\end{abstract}

\section{Time-dependent approaches to quantum problems in nuclear physics}
\label{sec:intro}
In this paper, we will report several applications of the real-time
calculations.
Although we concentrate our discussion on the nuclear physics problems,
the real-time approaches are useful in many fields of
many-body quantum systems.
Before discussing applications,
let us start from giving some reasons why we adopt the real-time calculations,
instead of a more prevalent time-independent method.

\subsection{Stationary solutions}

It is customary to solve a quantum mechanical problem in a form of
stationary equations, such as the eigenvalue problem of the
Schr\"odinger equation:
\begin{equation}
\label{Schroedinger_eq}
H\ket{\Phi_n}=E_n\ket{\Phi_n} .
\end{equation}
If a direct solution of this problem is achieved,
the solution provides a wave function of each energy eigenstate.
In principle, we may calculate all the physical observables using these
wave functions.
Amount of the computational task is roughly proportional to number of
eigenstates $\ket{\Phi_n}$ to be calculated.
Thus, for instance, when we are interested in the
excitation spectra and/or strength functions in a wide range of energy,
it should be computationally very demanding.

\subsection{Time-dependent solutions}

In contrast, a real-time solution of the time-dependent Schr\"odinger equation 
is, in general, a linear combination of energy eigenstates
$\ket{\Phi_n}$:
\begin{equation}
\label{td_state}
\ket{\Psi(t)}=\sum_n c_n e^{-i E_n t} \ket{\Phi_n} .
\end{equation}
The number of eigenstates and their coefficients $c_n$
involved in Eq.~(\ref{td_state}) are simply
determined by the initial state.
A trivial but important feature of the time-dependent solution,
Eq.~(\ref{td_state}), is that the state consists of
many eigenstates and distribution of their weights, $c_n$,
can be chosen as we wish, by selecting the initial wave packet.
Therefore, we may obtain physical quantities for states at a variety of
energies from a single time evolution of the wave packet,
using the Fourier decomposition into different energies.
Since the computational task does not strongly depend on the weights $c_n$,
the real-time calculation for the time-dependent equation
could be computationally more efficient than
solving the time-independent eigenvalue equation (\ref{Schroedinger_eq})
to obtain many eigenstates.

\subsection{Boundary condition for continuum states}

Another practical advantage of the real-time calculation
is the fact that the method does not require the boundary
condition for the description of the unbound continuum states.
The solution of Eq. (\ref{Schroedinger_eq}) for an unbound state at
positive energy $E_n$
cannot be uniquely determined without the asymptotic boundary condition.
In contrast, the solution of the time-dependent equation, Eq. (\ref{td_state}),
is unique for a given initial state.
In this sense, the boundary condition is naturally given by the
time-dependent solution.

To illustrate this fact,
let us consider scattering of two particles $A$ and $B$,
which are initially apart from each other, outside of the range of
interaction $V$.
These particles may be composite.
The Hamiltonian $H_0$ describes two free particles, $A$ and $B$,
at the initial channel.
The total Hamiltonian of the system is given by $H=H_0+V$.
The scattering solution for the energy eigenstate can be formally written as
\cite{Messiah}
\begin{equation}
\label{scatt_state_1}
\ket{\Psi^{(+)}(E)}=\left(1+\frac{1}{E-H+i\eta}V\right) \ket{\Phi(E)} ,
\end{equation}
where the initial (incident) state is an eigenstate of $H_0$,
$H_0\ket{\Phi(E)}=E\ket{\Phi(E)}$.
The Green's function, $(E-H+i\eta)^{-1}$,
contains the infinitesimal $i\eta$ that
determines the scattering boundary conditions.
The state of Eq. (\ref{scatt_state_1}) is obtained by
acting the M\o ller's wave operator on the initial state.
\begin{equation}
\label{scatt_state_2}
\ket{\Psi^{(+)}(E)}= U(0,-\infty) \ket{\Phi(E)} 
                   =\left( 1-\int_{-\infty}^0 dt' U(0,t')
                     e^{-\eta|t'|}V_I(t') \right) \ket{\Phi(E)} ,
\end{equation}
where $U(t,t')=e^{iH_0 t}e^{-iH(t-t')}e^{-iH_0 t'}$ and
$V_I(t)\equiv e^{iH_0 t}V e^{-iH_0 t}$.
Here, the initial state $\ket{\Phi(E)}$ is given by a plane wave for
the relative motion between $A$ and $B$.
Then, we need the convergence (adiabatic switch-off) factor $e^{-\eta|t'|}$,
in order to remove the interaction between $A$ and $B$
at the limit of $t\rightarrow -\infty$.
This trick is necessary when we treat the energy eigenstate,
because the particles $A$ and $B$ are overlapped in the plain-wave solution
and interacting each other.
In contrast, the infinitesimal factor $\eta$ becomes unnecessary \cite{Messiah}
if we use a wave packet, $\ket{\Phi_{\rm pac}}$,
instead of the eigenstate $\ket{\Phi(E)}$
in Eq. (\ref{scatt_state_2}).
Since the wave packet is spatially localized,
we can construct the initial state in which the wave packets of
$A$ and $B$ are far apart,
not interacting each other ($V\ket{\Phi_{\rm pac}}=0$).
The scattering wave packet,
which is a superposition of many eigenstates,
can be calculated by propagating the initial wave packet.
We may stop the time evolution at a finite period of time,
because the interaction takes place only in the finite time.
Once the wave packet is scattered away outside of the interacting region,
the state is governed by the Hamiltonian $H_0$.
After the scattering,
the proper scattering asymptotic behavior should be automatically imposed
for the wave packet.

\subsection{Intuitive picture of quantum dynamics}

The time-dependent description has not only
these advantages in practical computation, but also provides an
intuitive picture of the quantum dynamics.
It is very helpful for our insight into complex systems
to visualize movement of the wave packet.
This will be demonstrated in the following applications.

\subsection{Nuclear many-body problems}

The nucleus is a self-bound quantum system which presents a rich variety
of phenomena.
It is composed of fermions of spin $1/2$ and isospin $1/2$,
called nucleons (protons and neutrons),
interacting with each other through
a complex interaction with a short-range repulsive core \cite{BM69}.
Thus, the nucleus is a strongly correlated system.
Furthermore, it is highly ``quantum'', which could be quantified by
the magnitude of
the zero-point kinetic energy relative to the potential energy,
$\langle T \rangle /\langle V \rangle \approx 1$.
This is roughly unity in nuclear systems:
$\langle T \rangle /\langle V \rangle \approx 1$,
which indicates that the nucleus has a very strong quantum nature.
Even for the liquid helium, this ratio is smaller than the nuclear case.

Remarkable experimental progress in production and study of exotic nuclei
requires us to construct a theoretical model with higher accuracy and
reliability.
Extensive studies have been made in the past,
to introduce models and effective interactions to describe
a variety of nuclear phenomena and to understand basic nuclear dynamics
behind them \cite{BM69,RS80}.
Simultaneously, significant efforts have been made in the microscopic
foundation of those models.
For light nuclei,
the ``first-principles'' large-scale computation,
starting from the bare nucleon-nucleon (two-body \& three-body) forces,
is becoming a current trend in theoretical nuclear physics.
It is computationally very challenging, because, as we mentioned above,
the nucleus is a highly quantum, strongly correlated Fermionic many-body
system, interacting via the complex and ``singular'' nuclear force.
Although these ab-initio-type approaches
are still limited to nuclei with very small mass number,
they have recently shown a significant progress.
These issues are addressed by other contributions
\cite{Carlson,Quaglioni,Hagen}.

In this paper, we show a few nuclear-physics problems:
The low-energy nuclear fusion reactions, 
the nuclear response in the continuum, and
the time-dependent density-functional theory in the linear regime.
Here, we would like to emphasize again that,
although a particular model/theory is adopted for each calculation,
the concept of the real-time method is quite general and applicable
to other models and to other subfields of physics as well.

\section{Real-time calculation of sub-barrier fusion reaction}

In this section, we present the time-dependent wave-packet method
and discuss tunneling dynamics in sub-barrier fusion process
\cite{Yab97,YUN03-P,NYIKU04-P,IYNU06,IYNU07-P}.
To illustrate the essential idea, we start from
a simple two-body problem.

\subsection{Two-body time-dependent wave-packet model}
\label{sec:fusion_simple_model}.

A simple potential model of the fusion reaction is constructed
as follows:
The Hamiltonian for the relative motion between two nuclei
is given by ($\hbar=1$)
\begin{equation}
H=-\frac{1}{2\mu}\nabla_R^2 + V(R) +iW(R)
 =-\frac{1}{2\mu}
  \left(\frac{1}{R}\frac{\partial}{\partial R}R\right)^2
  +\frac{\vec{L}^2}{2\mu R^2}
  + V(R) +iW(R) .
\end{equation}
This leads to the radial Schr\"odinger equation in each partial wave:
\begin{equation}
\label{TDSE_1}
i\frac{\partial}{\partial t} u_L(R,t)
=
\left\{
-\frac{1}{2\mu}\frac{d^2}{dR^2}+\frac{L(L+1)}{2\mu R^2} + V(R) + iW(R)
\right\} u_L(R,t) ,
\end{equation}
where $\psi(R,t)=\sum_{LM}(u_L(R,t)/r )Y_{LM}(\hat{R})$.
The real potential, $V(R)$, consists of the repulsive Coulomb and
attractive nuclear potentials.
When the approaching two nuclei pass through the Coulomb barrier,
this is regarded as the fusion.
To calculate its probability,
it is convenient to use the imaginary potential,
$iW(R)$, that is non-zero only inside the Coulomb barrier.
Then, we define the fusion by the flux loss caused by $iW(R)$.
We assume here the collision of $^{10}$Be ($Z=4$) on $^{208}$Pb ($Z=82$).
$W(R)$ and the nuclear part of $V(R)$ are assumed to be in the
Woods-Saxon form.

\begin{figure}
\begin{minipage}{0.5\textwidth}
  \includegraphics[width=0.95\textwidth]{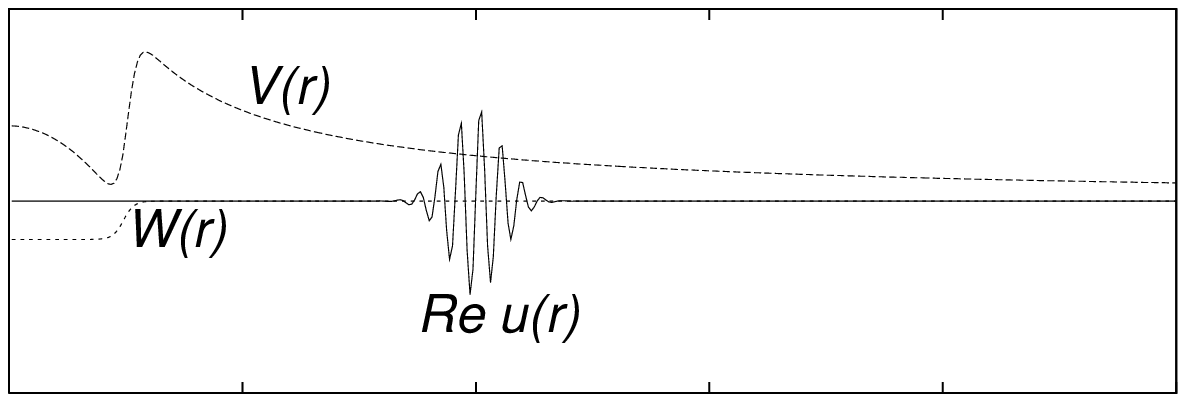}\newline
  \includegraphics[width=0.95\textwidth]{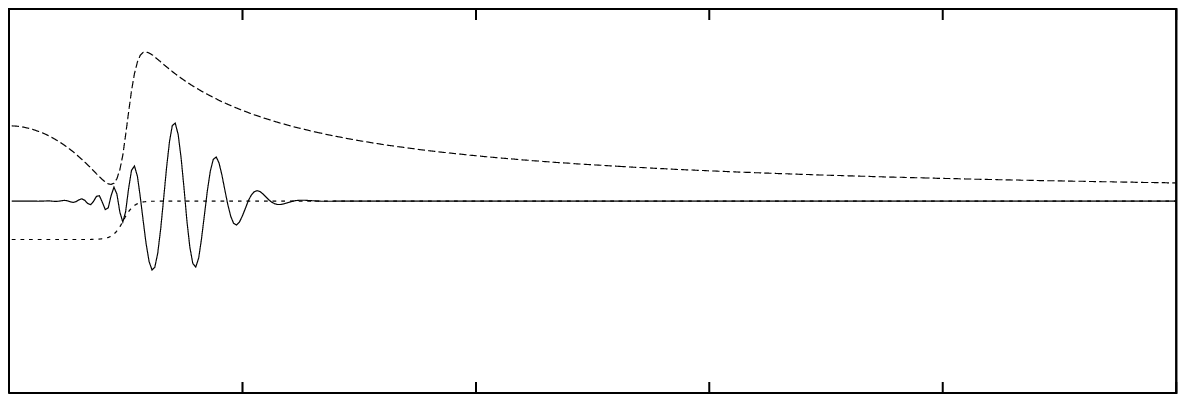}\newline
  \includegraphics[width=0.95\textwidth]{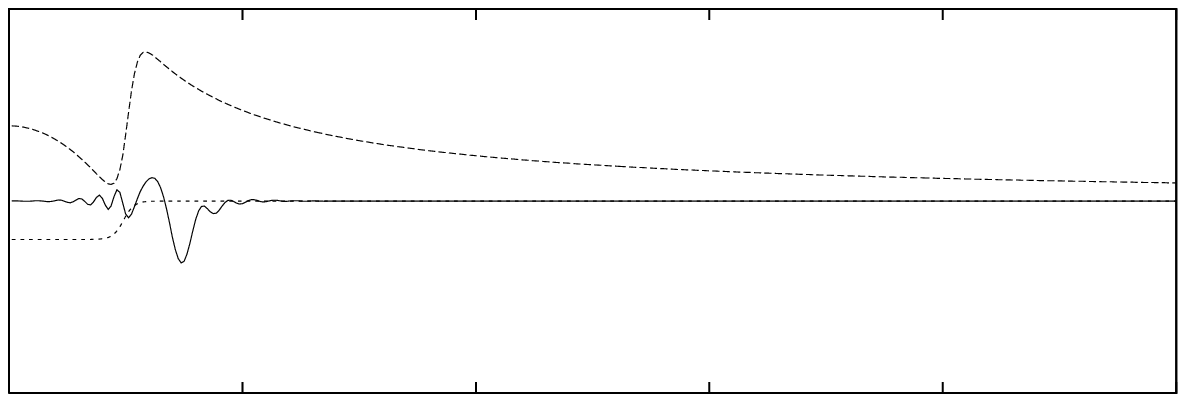}\newline
  \includegraphics[width=0.95\textwidth]{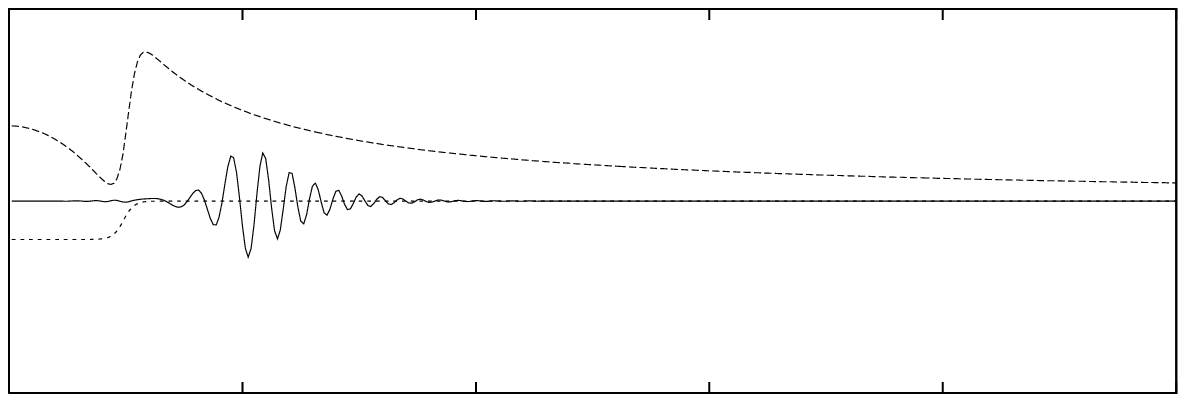}\newline
  \includegraphics[width=0.95\textwidth]{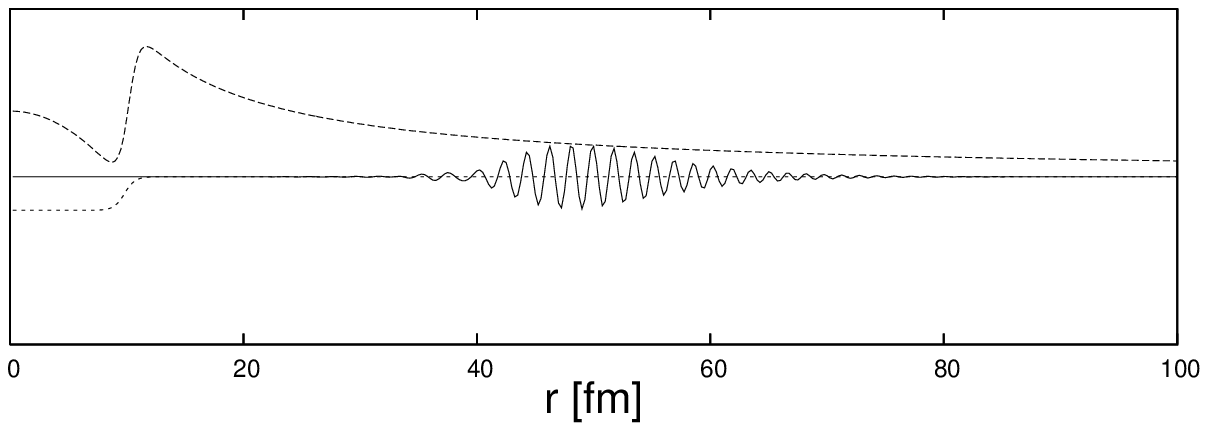}
\caption{Snap shots of time-dependent wave functions, ${\rm Re}\ u_{L=0}(r)$.
The panels are ordered in time from the top to the bottom.
The dashed (dotted) line indicates a shape of the real (imaginary)
part of the potential.
See text for details.
}
\label{fig:fusion_snap_2b}       
\end{minipage}
\hfill
\begin{minipage}{0.45\textwidth}
\rightline{
  \includegraphics[width=\textwidth]{figy2.eps}
}

\bigskip

\caption{Top panel: Energy distribution of the initial (dashed) and final
wave packets (solid line) corresponding to the top and bottom panels
in Fig.~\ref{fig:fusion_snap_2b}, respectively.\newline
Bottom: Calculated fusion probability ($L=0$) as a function of energy.
The solid line is calculated with the time-dependent wave-packet method,
while the circles indicate results obtained from
time-independent (fixed energy) solutions of the Schr\"odinger equation.}
\label{fig:fusion_prob_2b}       
\end{minipage}
\end{figure}

The most common way to solve this problem is to find
a stationary solution with a fixed energy $E$ by integrating
the time-independent radial Schr\"odinger equation.
Then, in the asymptotic region, one can compare the flux of
incoming and outgoing Coulomb waves,
$u_L(R)\sim C_L^{\rm in} u_L^{(-)}(R) + C_L^{\rm out} u_L^{(+)}(R)$,
to obtain the fusion probability,
$P_L = (|C_L^{\rm in}|^2-|C_L^{\rm out}|^2)/|C_L^{\rm in}|^2$.
Results obtained in this way are shown by circles in the bottom
panel of Fig~\ref{fig:fusion_prob_2b} for $L=0$.

We demonstrate an alternative approach: the time-dependent wave-packet method.
The initial wave function is of the Gaussian form
\begin{equation}
u_L(R,t=0)=e^{-iKR} e^{-\gamma^2 (R-R_0)^2} ,
\end{equation}
where $R_0$ indicate the initial position of the Gaussian center.
The parameters, $K$ and $\gamma$, are arbitrary, but
should be chosen so as to cover the interested energy region.
In the top panel of Fig.~\ref{fig:fusion_snap_2b},
the solid line shows the initial wave
packet for $L=0$,
the head-on collision with average relative momentum $K$.
In the present calculation, we set $K^2/2\mu = 28$ MeV and calculate
the time evolution up to $t_f=10$ MeV$^{-1}\approx 2000$ fm/c.
Time evolution of the wave packet is calculated by recursive operation
of the small-time evolution operator, $U(\Delta t)=e^{-i\Delta t H}$,
that is simply approximated with the fourth-order Taylor expansion,
$U(\Delta t)\approx \sum_{n=0}^4 (-i \Delta t H)^n/n!$.
The wave packet collides with the Coulomb barrier
in the second and third panels in Fig.~\ref{fig:fusion_snap_2b}.
Then, the wave packet is partially reflected back.
In the bottom panel,
the final wave packet contains only outgoing waves.
The missing part of the wave packet passes through the barrier,
entering the inner region and disappearing because of the absorbing
imaginary potential, $iW(R)$.
One can see, in the second and third panels,
that a part of the wave packet
actually penetrates the Coulomb barrier.
This incoming component vanishes
in the fourth and fifth panels.

Now, let us show the energy profile of the initial and final
wave packets.
We calculate the energy distribution of the wave packet at time $t$.
\begin{equation}
f_L(E;t)=\bra{u_L(t)}\delta(E-H)\ket{u_L(t)} .
\end{equation}
This quantity can be also calculated
using the time propagation technique \cite{Yab97}.
The dashed line in the top panel of Fig.~\ref{fig:fusion_prob_2b}
indicates the calculated energy distribution of the
initial wave packet, $f_{L=0}(E,t_i)$.
This is in a Gaussian form whose centroid is about 39 MeV that
corresponds to the given kinetic energy of 28 MeV.
The difference of 11 MeV is
the Coulomb potential energy at $R_0\sim 40$ fm.
The final distribution, $f_0(E,t_f)$ shown in the solid line in
Fig.~\ref{fig:fusion_prob_2b} (top panel), is calculated from the wave packet at
the bottom of Fig.~\ref{fig:fusion_snap_2b}.
We have $f_0(E,t_f)\approx 0$ for $E>42$ MeV.
These high energy parts end up the fusion.
On the other hand, the flux remains almost invariant
($f_0(E,t_f)\approx f_0(E,t_i)$) for $E<36$ MeV,
which suggests that the fusion cross section is negligible
in this low energy region.
These arguments can be quantified by calculating the fusion probability,
\begin{equation}
P_L(E)=\frac{f_L(E,t_i)-f_L(E,t_f)}{f_L(E,t_i)} .
\end{equation}
This is shown in the bottom panel of Fig.~\ref{fig:fusion_prob_2b}
by solid line.
The result calculated with the time-dependent wave-packet method
perfectly agrees with that of the fixed-energy calculation (circles).
This means that the time-dependent method provides an alternative
method of accurate calculation for fusion cross section.

\subsection{Three-body time-dependent wave-packet model}
\label{sec:fusion_three_body_model}
Let us upgrade the model to three bodies and discuss effects of
a valence neutron on the fusion.
There are many theoretical and experimental investigations
 on this, but the conclusion
is somewhat elusive yet.
Especially, the effect of weakly-bound valence neutrons with a spatially
extended wave function, which is often called ``neutron halo'',
is controversial.
There were many arguments that the weakly-bound neutron enhances
the fusion cross section, especially at sub-barrier energies
\cite{TS91,Hus92,TKS93,Hag00,DT02}.
However, as is shown in the followings, we
have reached the opposite conclusion \cite{Yab97,YUN03-P,NYIKU04-P,IYNU06,IYNU07-P}.

We assume that the projectile is a bound state of a neutron (n) and
a core nucleus (C), and the target (T) is treated as a point particle.
In this work, we neglect the intrinsic spin of neutron.
The time-dependent Schr\"odinger equation of the three-body
scattering is given as
\begin{equation}
i\hbar \frac{\partial}{\partial t} \Psi({\bf R},{\bf r},t)
= \left\{ -\frac{1}{2\mu} \nabla_{\bf R}^2
  -\frac{1}{2m}   \nabla_{\bf r}^2
  +V_{\rm nC}(r) +V_{\rm CT}(R_{\rm CT}) +V_{\rm nT}(r_{\rm nT}) \right\}
\Psi({\bf R},{\bf r},t),
\end{equation}
where we denote the relative n-C coordinate as ${\bf r}$ and the
relative P-T coordinate as ${\bf R}$.
The reduced masses of
n-C and P-T motions are $m$ and $\mu$, respectively.
The n-C potential, $V_{\rm nC}(r)$, is assumed to be real with the
Woods-Saxon form.
This potential must produce a bound state of the neutron around the core.
The depth of the potential $V_{\rm nC}$ is
varied according to the orbital energy
simulating a projectile nucleus with a tightly-bound to weakly-bound neutron.
The core-target potential, $V_{\rm CT}(R_{\rm CT})$,
contains an imaginary part $iW(R_{\rm CT})$ that simulates the fusion.
The n-T potential, $V_{\rm nT}(r_{\rm nT})$, is taken to be real.
The imaginary potential is present only between the core and target
nuclei.
When the core and target become close enough, the wave function will
vanish, to be counted as fusion.
Therefore, since the final destination of the neutron is irrelevant,
the present calculation does not distinguish complete
and incomplete fusion.
The total fusion probability (sum of complete and incomplete ones)
is calculated below.

The wave-packet is constructed using the partial wave expansion
in the body-fixed frame \cite{Pack74,Ima87}.
This calculation is equivalent to the one in the space-fixed frame.
However, the body-fixed frame has an advantage for calculations of
states with non-zero total angular momentum $J\neq 0$
($\vec{J}=\vec{L}+\vec{l}$).
In the body-fixed frame, channels are characterized by
the magnetic quantum number $\Omega$ (the projection of $J$ on the $z$-axis)
and the angular momentum $l$ conjugate to the angle $\theta$ between
$\mathbf{r}$ and $\mathbf{R}$.
Coupling between different $\Omega$ channels is present only for
$\Delta\Omega=\pm 1$, caused by the Coriolis term.
In the present paper, we focus our discussion on the head-on collision
($J=0$).
See Ref. \cite{IYNU06} for the complete calculation.
The $J=0$ state is characterized by $\Omega=0$ and $L=l$,
given by
\begin{equation}
\label{l_max}
\psi^{J=0}(R,r,\theta;t)
=
\sum_{l=0}^{l_{\rm max}} \frac{u_{l}(R,r,t)}{Rr} 
\frac{\sqrt{2l+1}}{4\pi} P_l(\cos\theta) .
\end{equation}
Here, we need to truncate the space by setting
the maximum value of the partial waves $l$.
To calculate the Coulomb breakup process, we may take
$l_{\rm max}=4$.
However, when the n-T potential is included,
nuclear breakup and the neutron transfer may take place, then,
much larger $l_{\rm max}$ is necessary to obtain convergent results;
$l_{\rm max}\sim 70$ \cite{YUN03-P,IYNU06}.
Since the present calculation is simply based on
the discretization of the radial coordinate,
it is straightforward to include these high partial waves.
This is an advantage over
the continuum-discretized coupled-channel calculations \cite{DT02}.

In this model, the projectile is assumed to have a simple structure that
a neutron sits on the 2$s$ orbital in the n-C potential $V_{\rm nC}$:
$\phi_0({\bf r})=\sqrt{1/4\pi} y_0(r)/r$.
The initial wave packet at time $t=0$ is prepared
in the same manner as we have shown in Sect.~\ref{sec:fusion_simple_model}.
Namely, the P-T relative wave function is given by
the boosted Gaussian wave packet whose centroid is at $R_0=25$ fm.
\begin{equation}
u_l(R,r,t=0) =
e^{-iK_0 R} e^{-\gamma^2 (R-R_0)^2} y_0(r) \delta_{l0} .
\end{equation}

\begin{figure}
\centerline{
  \includegraphics[width=.5\textwidth]{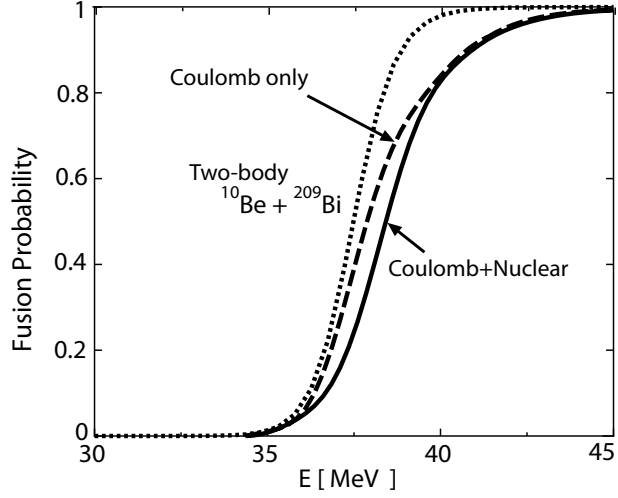}
}
  \caption{
  Fusion probability ($J=0$) as a function of incident energy for
  halo $^{11}$Be ($\epsilon_n=-0.6$ MeV) on $^{209}$Bi.
  Dotted line indicates the probability for the projectile without
  the valence neutron.
  Dashed and solid lines indicate fusion probabilities
  calculated with and without $V_{\rm nT}$ potentials,
  respectively.
  }
  \label{fig:fusion_prob_3b}
\end{figure}

\subsubsection{Case (1) Well-bound projectile: }
First, we investigate fusion reaction of projectile with a
well-bound neutron.
For this case, we found that the fusion cross section is identical
to the one without the valence neutron.
Thus, the valence neutron essentially sticks to the core nucleus,
and they move together as a single nucleus.
This may be naturally expected.

However, there are exceptions.
when the condition of energy matching is satisfied,
which means that energies of the neutron orbital in the projectile
and target are approximately degenerate,
we observe a substantial enhancement of sub-barrier fusion cross section.
The fusion probability is strongly correlated with the
neutron transfer probability and sensitive to $V_{\rm nT}$.
This may be understood in terms of adiabatic dynamics of
the valence neutron.
See discussion in Ref.~\cite{Yab97}.

\begin{figure}
  \includegraphics[height=.19\textwidth,angle=90]{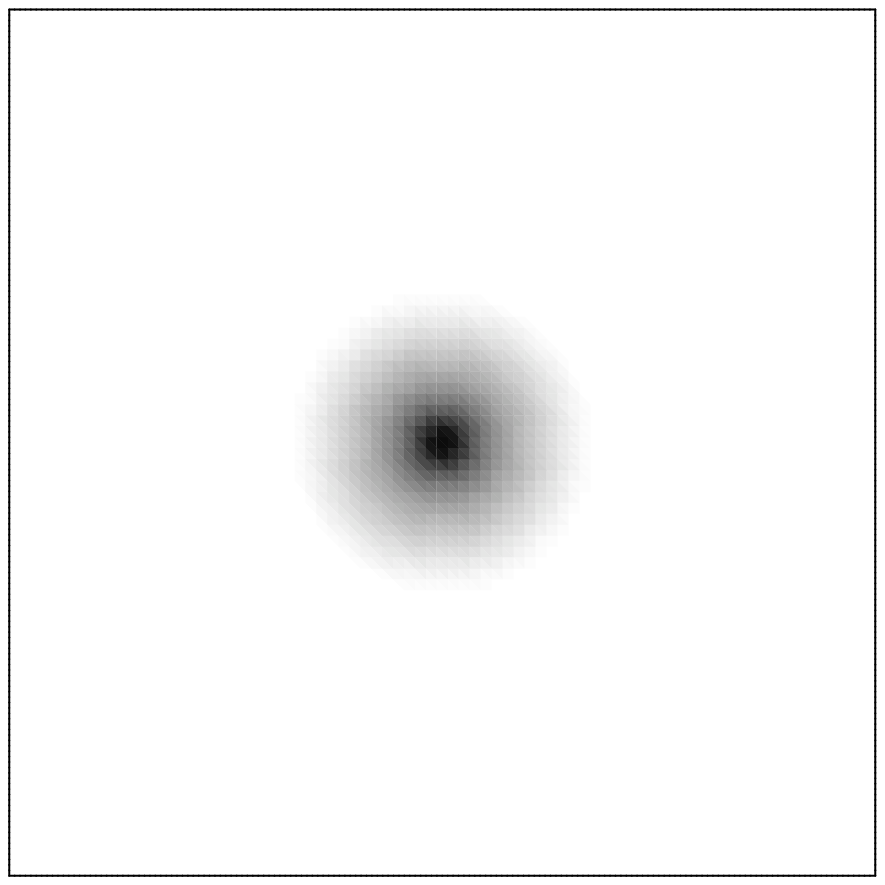}
  \includegraphics[height=.19\textwidth,angle=90]{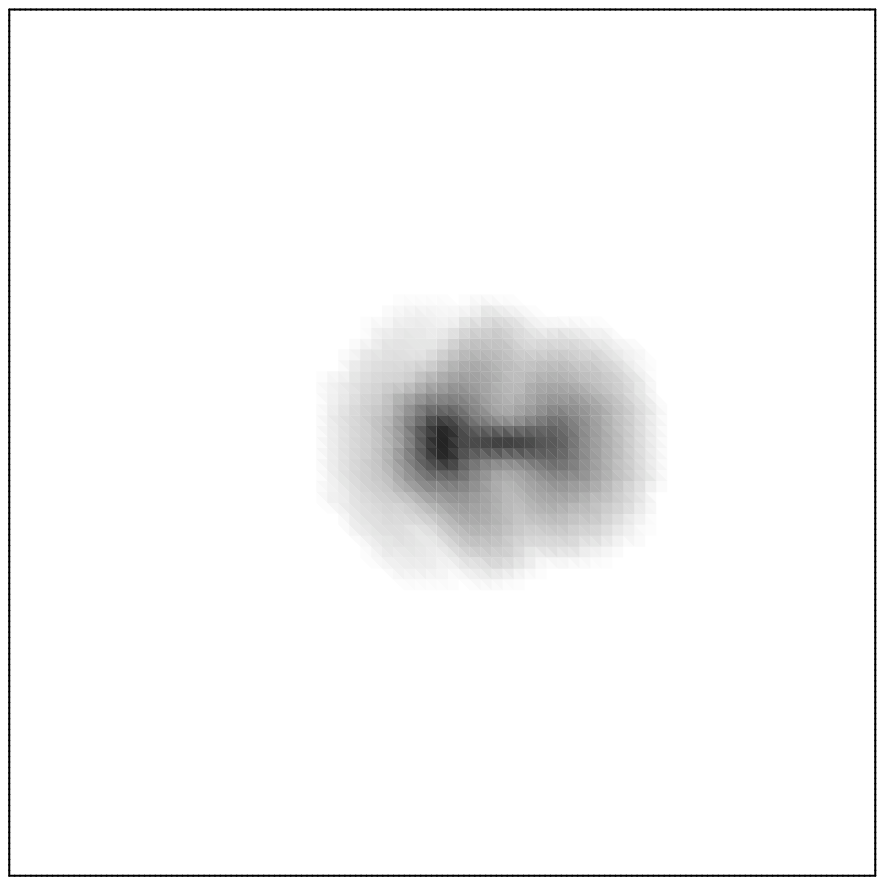}
  \includegraphics[height=.19\textwidth,angle=90]{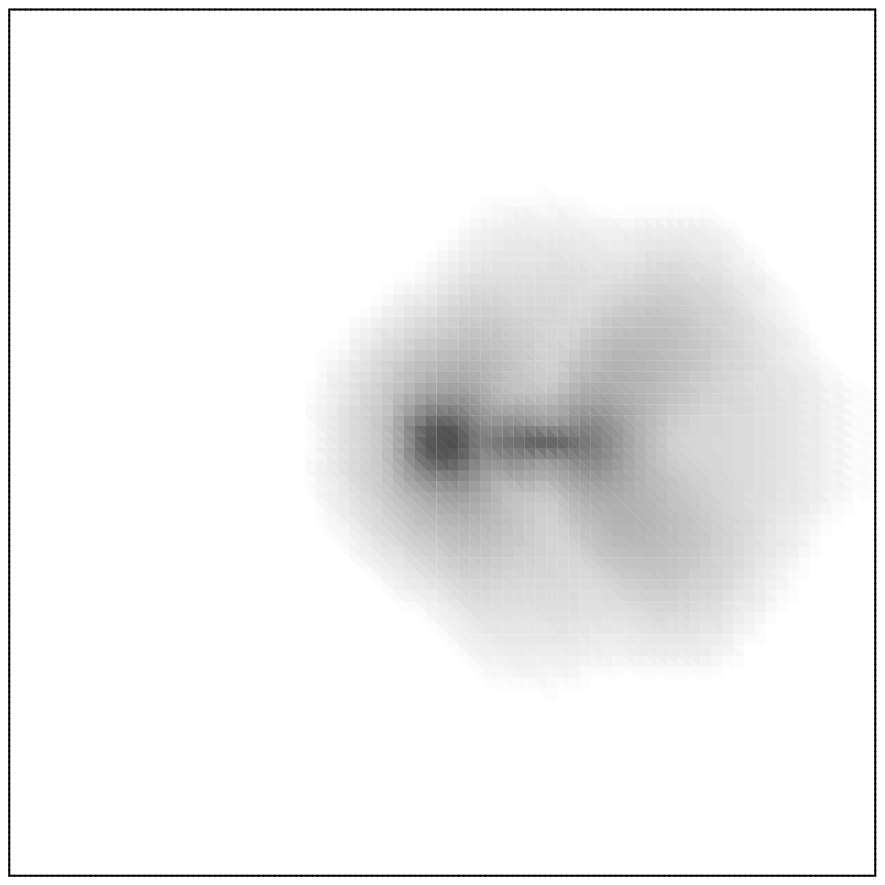}
  \includegraphics[height=.19\textwidth,angle=90]{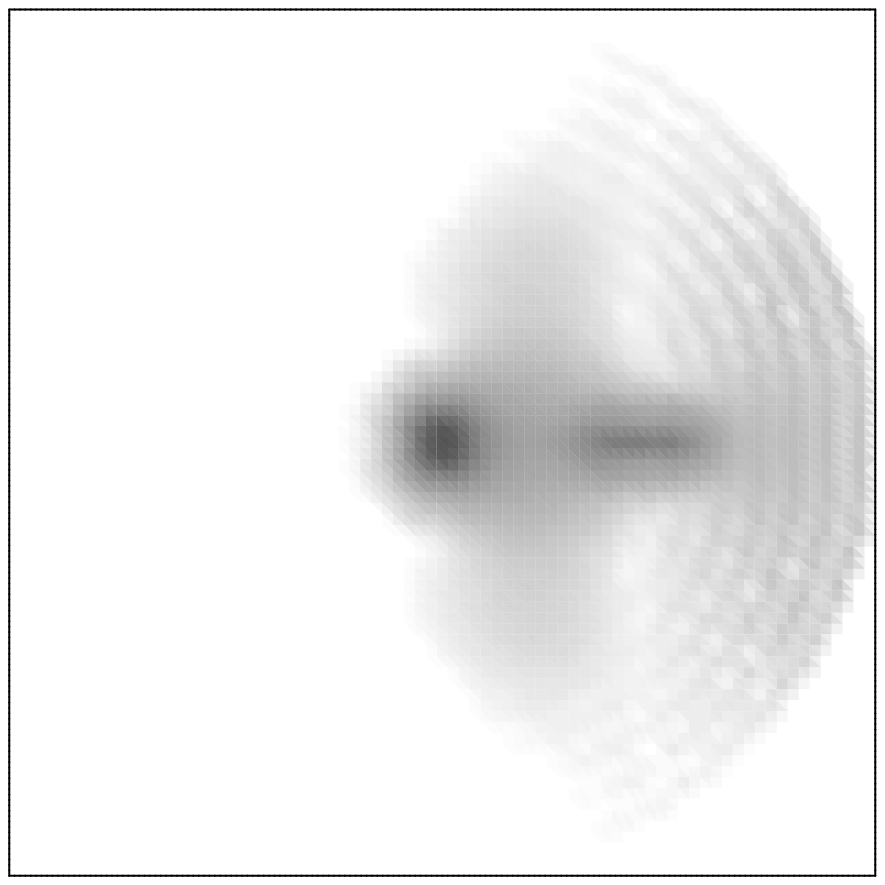}
  \includegraphics[height=.19\textwidth,angle=90]{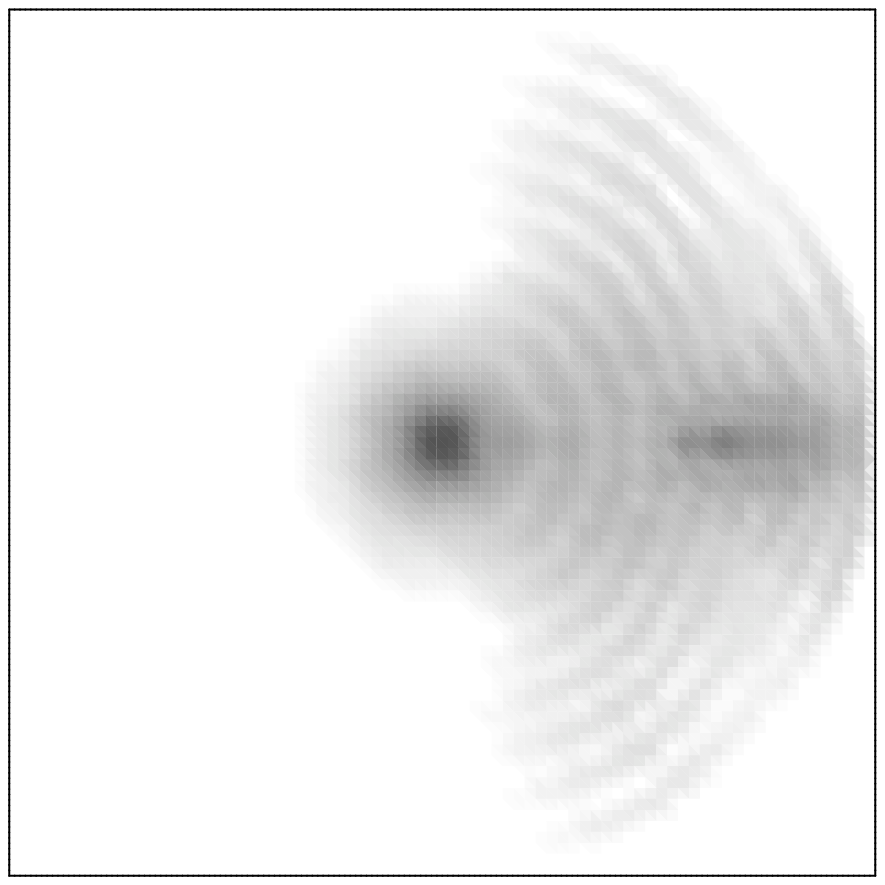}
  \caption{
  Time-dependent density distribution of the valence neutron,
  $\rho(r,\theta;t)$ in the collision process of $^{11}$Be on $^{209}$Bi.
  The panel at the left (right) end shows the initial (final) wave packet.
  The center of each panel, $r=0$, indicates the position of the core,
  and the right is the direction for the target, $\theta=0$.
  }
  \label{fig:fusion_snap_3b}
\end{figure}

\subsubsection{Case (2) Weakly-bound projectile (Neutron halo): }
Now, let us discuss fusion of weakly bound $^{11}$Be on $^{209}$Bi,
namely effects of the neutron halo.
As is mentioned above,
there are theoretical arguments that the weakly-bound neutron enhances
the sub-barrier fusion cross section \cite{TS91,Hus92,TKS93,Hag00,DT02}.
We studied the same problem in a fully microscopic framework with
the real-time method.

Calculated fusion probabilities at the head-on collision ($J=0$)
are shown in Fig.~\ref{fig:fusion_prob_3b}.
The fusion probability is slightly suppressed by the presence of
the halo neutron.
It seems that the fusion dynamics for the neutron-halo projectile
are mainly determined by the Coulomb breakup process.
Time evolution of the neutron density distribution during the
collision process is shown in Fig.~\ref{fig:fusion_snap_3b}.
We plot the following quantity in the $x$-$y$ plane
with $x=r\cos\theta$ and $y=r\sin\theta$
for the head-on collision of $J=0$;
\begin{equation}
\rho(r,\theta;t)=\int dR |\psi^{J=0}(R,r,\theta;t)|^2 .
\end{equation}
The direction of the target always corresponds to $\theta=0$,
which means that the target nucleus approaches from the right side
on the $x$-axis, then returns back to the right.
Figure~\ref{fig:fusion_snap_3b} suggests the followings:
When the core nucleus is decelerated by the target Coulomb field,
the halo neutron behaves like a spectator, keeping its incident velocity and
leaving the core nucleus.
This yields the Coulomb breakup of the projectile, then
reduces the collision energy between the core and the target.
This spectator picture of the fusion dynamics for halo nuclei
nicely explains why the total fusion cross section does not increase
even though the radius of the neutron-halo nucleus is so large
\cite{Sig04}.

Our conclusion is that the presence of the halo neutron suppresses 
the fusion cross section, irrelevant to the collision energy.
This contradicts other former predictions
\cite{TS91,Hus92,TKS93,Hag00,DT02}.
In fact, the model adopted in the coupled-channel calculation
in Ref.~\cite{DT02} is very close to ours.
We have found that the difference comes from the truncation of the
model space \cite{IYNU06}.
Namely, the result in Ref.~\cite{DT02} did not reach the convergence
with respect to the number of the partial wave $l_{\rm max}$ in
Eq. (\ref{l_max}).
If we set $l_{\rm max} \leq 4$ as is done in Ref.~\cite{DT02},
we also obtain an enhanced fusion cross section at sub-barrier energies.
However, the cross section decreases as $l_{\rm max}$ increases, and
finally it becomes even smaller than the two-body result
without the valence neutron.

\section{Real-time calculation of strength functions in the continuum}

In this section, we discuss a method of calculating the strength
distribution, in particular, 
the $E1$ strength distribution, from the time-dependent wave function.

The $E1$ strength distribution in the continuum is defined by
\begin{equation}
\frac{dB(E1;E)}{dE}\equiv 
\sum_{m,m'} \int dE' \delta(E-E') 
\left|\bra{\Phi^{(+)}(E';m')} M(E1;m) \ket{\Phi_0} \right|^2.
\label{E1}
\end{equation}
Here, $M(E1;m)$ is the electric dipole operator with the magnetic
quantum number $m$.
The final states $\ket{\Phi^{(+)}(E;m)}$ should have
the proper outgoing asymptotic
form when unbound channels are open.
The construction of the final state in the continuum is 
a difficult task in general, especially for many-body non-spherical
systems.
An alternative way of avoiding this difficulty 
is the time-dependent description.
Let us show a simple example as an illustration of the method.

\subsection{Illustrative example}

The $E1$ strength function of Eq. (\ref{E1})
can be written in a time-dependent form:
\begin{equation}
\frac{dB(E1;E)}{dE} =
-\frac{1}{\pi}{\rm Im} \sum_{m} i \int_0^\infty dt e^{i(E+i\eta)t}
\bra{\Phi_0} M^\dagger(E1;m) e^{-itH} M(E1;m)\ket{\Phi_0} .
\label{TD_E1}
\end{equation}
This is easily obtained by the fact that
the time integration simply produces the Green's function $(E-H+i\eta)^{-1}$
whose imaginary part is nothing but the delta function $\delta (E-H)$.
Therefore, we can calculate the $E1$ strength distribution as follows:
(1) Construct the initial state by operating the $E1$ operator to the
ground state, $\ket{\Psi(t=0)}=M(E1;m)\ket{\Phi_0}$.
(2) Calculate its time evolution $\ket{\Psi(t)}$ and
its overlap with the initial state.
(3) Its Fourier transform leads to $dB(E1;E)/dE$.
In this time-dependent formulation, we do not need to construct
the final states with the proper boundary condition.
This is automatically taken into account in the time evolution of
wave packets.
When the $E1$ operator excites the ground state into unbound continuum states,
the wave packet $\ket{\Psi(t)}$ will decays in time and
the integrand of Eq.~(\ref{TD_E1}) will vanish at $t\rightarrow\infty$.
Thus, we do not need the convergence factor $+i\eta$.

\begin{figure}
\begin{minipage}[c]{0.45\textwidth}
  \includegraphics[width=0.9\textwidth]{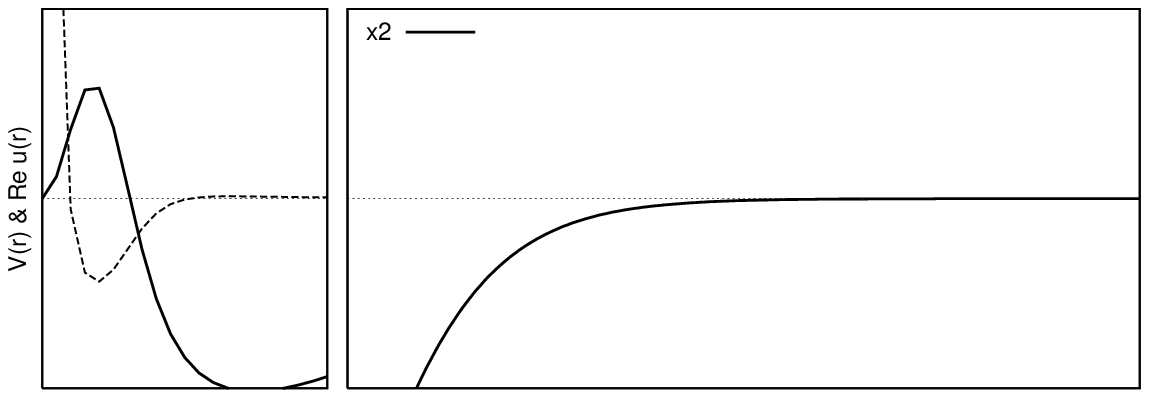}\newline
  \includegraphics[width=0.9\textwidth]{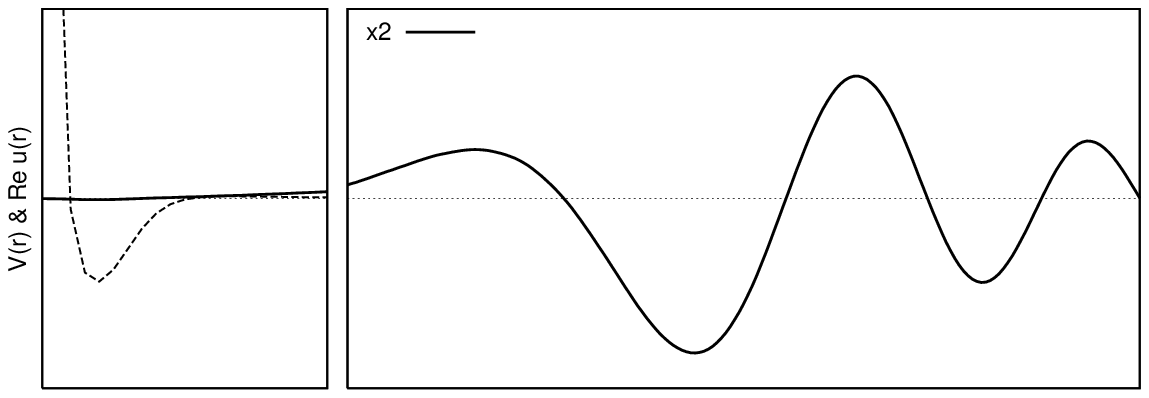}\newline
  \includegraphics[width=0.9\textwidth]{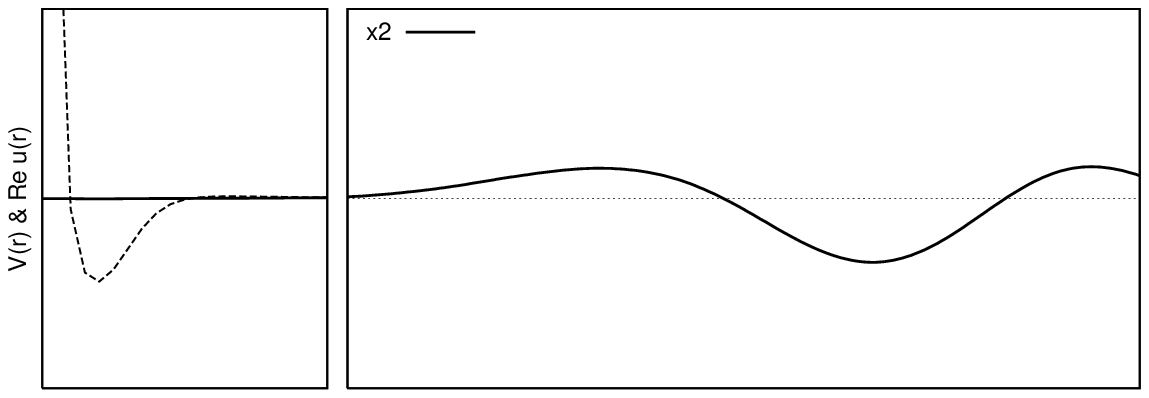}\newline
  \includegraphics[width=0.9\textwidth]{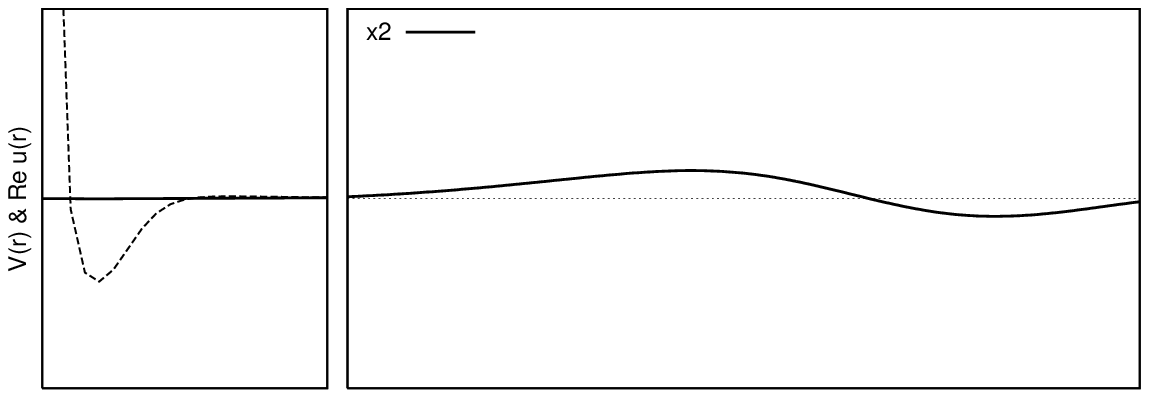}\newline
  \includegraphics[width=0.9\textwidth]{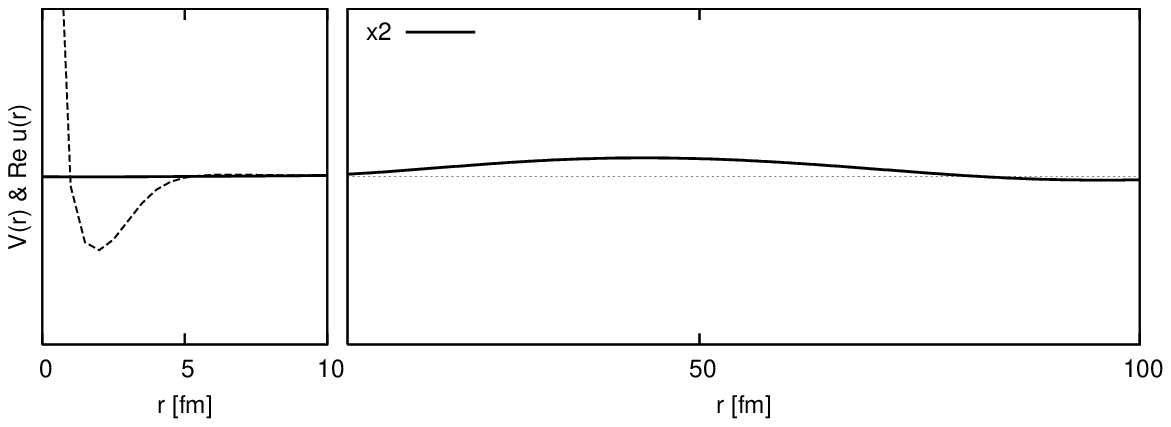}
  \caption{
  Time evolution of a state (real part of the wave function)
  initially excited by the dipole operator.
  The vertical scale for $r=10\sim 100$ fm is magnified by a factor of
  two compared to that for $r=0\sim 10$ fm.
  The dashed line indicate the shape of the $p$-wave potential.
  }
  \label{fig:response_snap_2b}
\end{minipage}
\hfill
\begin{minipage}[c]{0.45\textwidth}
  \includegraphics[width=\textwidth]{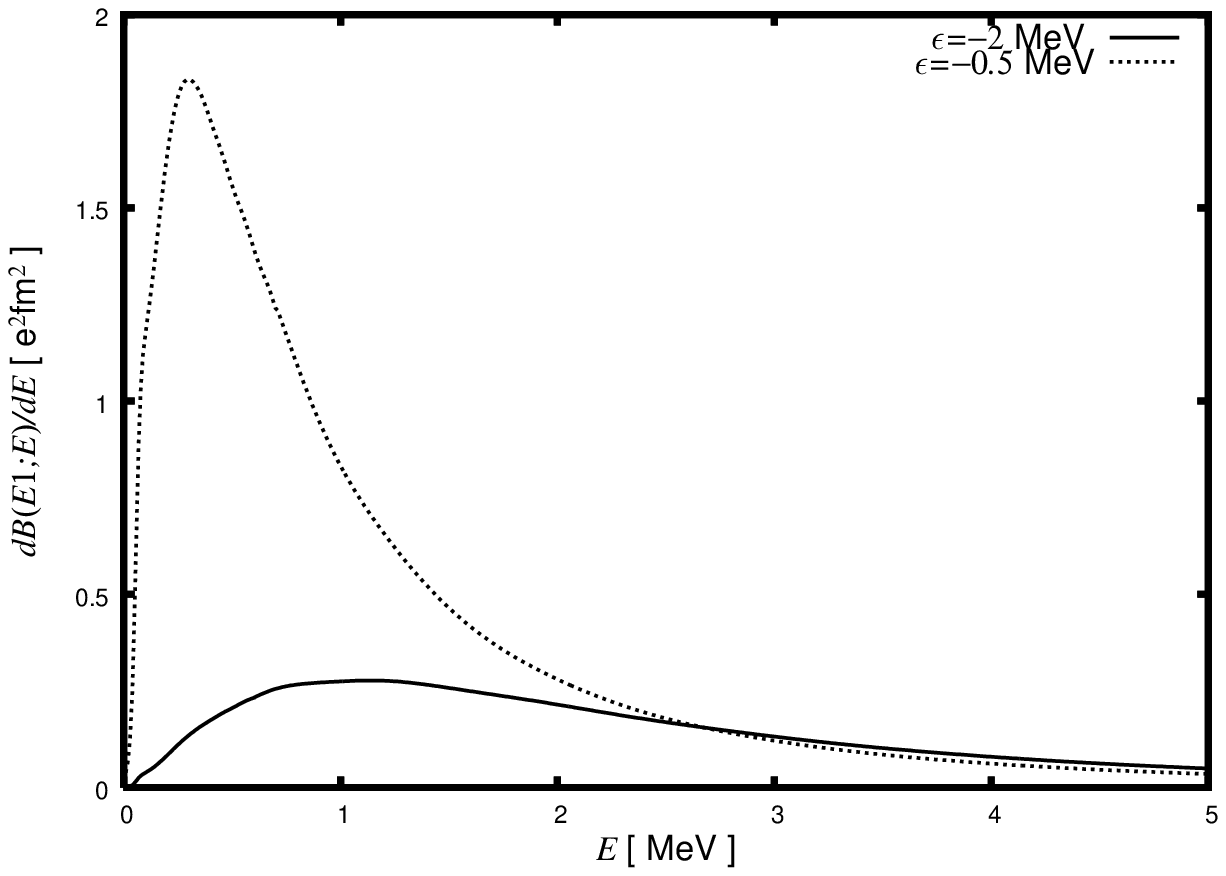}
\caption{Calculated $E1$ strength distribution for $^{11}$Be.
See text for explanation.
}
\label{fig:E1_2b}       

\end{minipage}
\end{figure}

Let us discuss application of the time-dependent method
to a simple single-particle model.
We again use the Woods-Saxon potential of $V_{\rm nC}$
in Sect.\ref{sec:fusion_three_body_model}
to describe the $^{11}$Be nucleus as a two-body system of
a $^{10}$Be core and a neutron.
The neutron is assumed to sit on the $2s$ orbital.
Multiplying it by the $E1$ operator with the recoil charge,
we construct the initial state
which is shown in the top panel of Fig.~\ref{fig:response_snap_2b}.
We solve the time-dependent Schr\"odinger equation similar to
Eq.~(\ref{TDSE_1}) with $L=1$.
The imaginary potential $iW(R)$ is active at $R>50$ fm
to absorb outgoing waves.
The time evolution is shown in the following panels
in Fig.~\ref{fig:response_snap_2b}.
The state with $L=1$ propagates into $r\rightarrow\infty$ in time.
After a certain period, the state no longer has an overlap matrix element
with the initial state.
Then, we stop the calculation.
From this time evolution, we obtain $E1$ strength distribution (dotted line)
shown in Fig.~\ref{fig:E1_2b}.
There is a strong low-energy peak known as the threshold effect.
This peak is drastically reduced if we change the energy of
the $2s$ orbital.
For instance, the solid line in Fig.~\ref{fig:E1_2b} shows the result
for the case that the energy of the $2s$ orbital is $-2$ MeV.
Experimental data \cite{Nak94,Pal03,Fuk04} indicate
a similar shape but somewhat smaller strength.

This method can be extended into many-body problems in a straightforward
manner.
However, its computational task drastically increases as the number of
particles increases.
Thus, we resort to the density functional model of nuclei in the
next section.

\label{sec:response_simple_model}

\begin{figure}
\end{figure}

\section{Density-functional approach to nuclei}

The density functional theory is a leading theory for describing
nuclear properties of heavy nuclei and perhaps the only theory capable of
describing all nuclei and nuclear matter with a single universal
energy density functional.
In the nuclear physics, it is often called the self-consistent mean-field
model, because of a historical development based on the
Brueckner-Hartree-Fock theory and introduction of the effective interaction.
However, it should be noted that there are both fundamental and practical
differences between
the naive mean-field theory and the mean-field model of the nucleus which
is, in fact, conceptually analogous to the density-functional theory.

I briefly review basic properties of nuclei and show that those
properties cannot be understood by a simple mean-field model.
Here, the saturation property plays a key role.

\subsection{Nuclear saturation in the mean-field model}
\label{sec:IPM}

Nuclei are known to be well characterized by the saturation property.
Namely, they have an approximately constant density
$\rho_0\approx 0.17$ fm$^{-3}$,
and a constant binding energy per particle $B/A\approx 16$ MeV.\footnote{
This is the extrapolated value for the infinite nuclear matter
without the surface and the Coulomb energy.
The observed values for finite nuclei are
$B/A\approx 8$ MeV.}
In this section, I show
that the nuclear saturation property has a great impact on
nuclear models.
Especially, it is inconsistent with the independent-particle model of nuclei
with a ``naive'' average (mean-field) potential.

There are many evidences for the fact that
the mean-free path of nucleons
is larger than the size of nucleus.
In fact, the mean free path depends on the nucleon's energy,
and becomes larger for lower energy \cite{BM69}.
Therefore, it is natural to assume that
the nucleus can be primarily approximated by the
independent-particle model with an average one-body potential.
The crudest approximation is the degenerate Fermi gas of the same number of
protons and neutrons ($Z=N=A/2$).
The observed saturation density of $\rho_0\approx 0.17$ fm$^{-3}$
gives the Fermi momentum, $k_F\approx 1.36$ fm$^{-1}$,
that leads to the Fermi energy (the maximum kinetic energy),
$T_F=k_F^2/2M\approx 40$ MeV.

First, I show that the independent-particle model
with a constant attractive potential $V<0$ cannot describe
the nuclear saturation property.
It follows from the simple arguments.
The constancy of $B/A$ means that it is approximately equal to the
separation energy of nucleons, $S$.
In the independent-particle model, it is estimated as
\begin{equation}
\label{B1}
S \approx B/A \approx -(T_F + V) .
\end{equation}
Since the binding energy is $B/A\approx 16$ MeV,
the potential $V$ is about $-55$ MeV.
It should be noted that the relatively small separation energy is
the consequence of the significant cancellation between
kinetic and potential energies.
The total (binding) energy is given by
\begin{equation}
\label{B2}
-B = \sum_{i=1}^A \left( T_i + \frac{V}{2} \right)
        = A \left( \frac{3}{5}T_F + \frac{V}{2} \right) ,
\end{equation}
where we assume that the average potential results from a two-body
interaction.
The two kinds of expressions for $B/A$, Eqs. (\ref{B1}) and (\ref{B2}),
lead to $T_F\approx -5V/4\approx 70$ MeV,
which is different from the previously estimated value ($\sim 40$ MeV).
Moreover, it contradicts the fact
that the nucleus is bound ($T_F < |V|$).

To reconcile
the independent-particle motion with the saturation property of the nucleus,
the nuclear average potential should be state dependent.
Allowing the potential $V_i$ depend on the state $i$,
the potential $V$ should be replaced by that for the highest occupied
orbital $V_F$ in Eq. (\ref{B1}),
and by its average value $\langle V \rangle$ in
the right-hand side of Eq. (\ref{B2}).
Then, we obtain the following relation:
\begin{equation}
\label{V_F}
V_F \approx \langle V \rangle + T_F/5 + B/A .
\end{equation}
Therefore, the potential $V_F$ is shallower 
than its average value.

Weisskopf suggested the momentum-dependent potential $V$, which can be
expressed in terms of an effective mass $m^*$ \cite{Wei57}:
\begin{equation}
\label{mom_dep_pot}
V_i=U_0+U_1\frac{k_i^2}{k_F^2} .
\end{equation}
Actually, if the mean-field potential is non-local, it can be
expressed by the momentum dependence.
Equation (\ref{mom_dep_pot})
leads to the effective mass, $m^*/m = (1+U_1/T_F)^{-1}$.
Using Eqs. (\ref{B1}), (\ref{V_F}), and (\ref{mom_dep_pot}),
we obtain the effective mass as
\begin{equation}
\label{m*/m}
\frac{m^*}{m} = \left\{ \frac{3}{2} + \frac{5}{2}\frac{B}{A}\frac{1}{T_F}
 \right\}^{-1} \approx 0.4 .
\end{equation}
Quantitatively, this value disagrees with the experimental data.
The empirical values of the effective mass
vary according to the energy of nucleons,
$0.7 \lesssim m^*/m \lesssim 1$,
however, they are almost twice larger than
the value in Eq. (\ref{m*/m}).
As far as we use a normal two-body interaction,
this discrepancy should be present in the mean-field calculation
with any interaction,
because Eq. (\ref{m*/m}) is valid in general
for a saturated self-bound system.
Therefore, the conventional models cannot simultaneously
reproduce the most basic properties of nuclei; the binding energy and
the single-particle property.

This problem can be solved in the density functional theory.
In nuclear physics, basically the same solution was interpreted
in terms of the density-dependent interaction.
The variation of the total energy density functional $E[\rho]$
with respect to the density
contains ``re-arrangement potential'',
$\partial V_{\rm eff}[\rho]/\partial\rho$,
which appear due to the density dependence of the effective force
$V_{\rm eff}[\rho]$.
These terms turn out to be crucial to obtain the saturation condition.
Now, the expression for the total energy, Eq. (\ref{B2}), should be modified
to include the re-arrangement effect.
This resolves the previous issue, then provides a consistent
independent-particle description for the nuclear saturation.
This is important to understand basic nuclear properties and to describe
nuclei in a universal framework.

Intensive studies in nuclear density functional models in recent years
have produced numerous results and new insights into nuclear structure
\cite{BHR03,LPT03}.
However, it is impossible to review all of them in this paper.
Thus, I will present results of our study
with the (time-dependent) density functional approach,
on the shape phase transition in the rare-earth nuclei and
on the photoabsorption cross section in $^{238}$U.

\subsection{Density-functional theory for superfluid nuclei}

A modern energy functional for nuclei is a functional of many kinds
of density, such as kinetic $\tau(\vec{r})$
and spin-orbit density $\vec{J}(\vec{r})$.
In addition, 
it is known that the heavy nuclei in the open-shell configurations
show characteristic features of the superfluid systems \cite{RS80}.
To describe superfluid properties produced by pairing correlation among
nucleons,
we need to add the pair (abnormal) density $\kappa(\vec{r})$.
These densities are collectively denoted as $\tilde{\rho}$ in the
followings.
Variation of the total energy, $E[\tilde{\rho}]$,
leads to the following equation, known as the Hartree-Fock-Bogoliubov equation
in nuclear physics \cite{RS80}:
\begin{equation}
\label{HFB}
H[\Psi,\Psi^*] \ket{\Psi_\mu} =
E_\mu \ket{\Psi_\mu} ,
\end{equation}
\begin{wrapfigure}{r}{7cm}
\includegraphics[width=7cm]{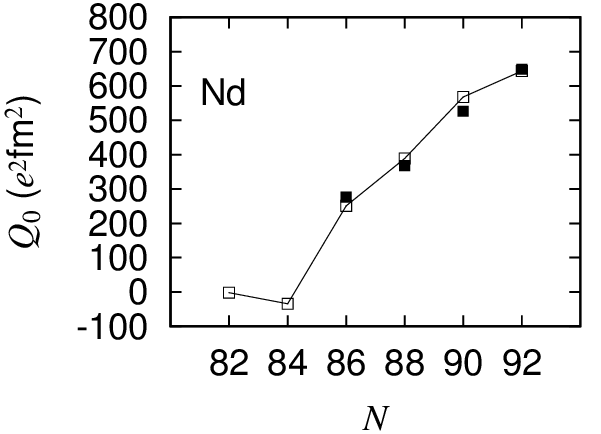}
\caption{Calculated (open squares) and experimental (filled)
intrinsic quadrupole moment for Nd isotopes \cite{YN11}.}
\label{fig:Nd_def}
\end{wrapfigure}
where $E_\mu$ and
$\ket{\Psi_\mu}$ are quasi-particle energies
and states, respectively \cite{RS80}.
$\ket{\Psi_\mu}$ is composed of two components;
the upper $\ket{U_\mu}$ and
the lower one $\ket{V_\mu}$.
The solution of Eq. (\ref{HFB}) defines the normal density
$\rho(\vec{r})=\sum_\mu V_\mu(\vec{r}) V_\mu^*(\vec{r})$,
the pair density
$\kappa(\vec{r})=\sum_\mu U_\mu(\vec{r}) V_\mu(\vec{r})$,
and other densities at the ground state.
Since $h[\Psi,\Psi^*]$ depends on these densities,
Eq. (\ref{HFB}) must be solved in a self-consistent way.
Minimization of the energy density functional may lead to
a spontaneous breaking of symmetry.
An example is given in Fig. \ref{fig:Nd_def} for Nd isotopes \cite{YN11}.
The intrinsic quadrupole moment calculated with the Skyrme
functional of SkM* is compared with the
experimental data.
At $N=82$, the nucleus at the ground state is spherical $Q_0=0$,
while for $N=86\sim 92$,
the deformation gradually develops.
The observed ground-state deformations deduced from the transition probability
$B(E2; 2^+ \rightarrow 0^+)$ are nicely reproduced.
Note that there are no adjustable parameters in this calculation.

\subsection{Time-dependent density-functional theory for superfluid nuclei}

For a description of the time-dependent phenomenon,
we need to extend the energy functional
to include the time-odd densities, such as the spin density $\vec{s}(\vec{r},t)$
and the current density $\vec{j}(\vec{r},t)$.
Now, all the densities are time dependent.
The time-dependent version of Eq. (\ref{HFB}) is formally written as
\begin{equation}
\label{TDHFB}
i\frac{\partial}{\partial t} \ket{\Psi_\mu(t)} =
H[\Psi(t), \Psi^*(t)] \ket{\Psi_\mu(t)} .
\end{equation}
This is known as the time-dependent Hartree-Fock-Bogoliubov equation
in nuclear physics.

The computation of Eq. (\ref{TDHFB}) is very demanding because we
need to calculate the time evolution of all the
quasi-particle states $\ket{\Psi_\mu}$.
The number of them is same as the dimension of the single-particle
model space which is much larger than the particle number.
Although we proposed a feasible approach to its linear regime,
known as the finite amplitude method (FAM) \cite{NIY07,INY09,INY11,AN11,Sto11},
the FAM is based on a time-independent formulation.
In this section, we present another approximate treatment of Eq. (\ref{TDHFB})
to provide time-dependent equations.

The approximate equations alternative to
Eq. (\ref{TDHFB}) is called the canonical-basis TDHFB \cite{Eba10}.
Assuming the diagonal property of the pair potential in the
canonical pair of states $k$ and $\bar k$,
Eq. (\ref{TDHFB}) can be approximated by
the following set of equations.
\begin{subequations}
\label{Cb-TDHFB}
\begin{eqnarray}
\label{dphi_dt}
&&i\frac{\partial}{\partial t} \ket{\phi_k(t)} =
(h(t)-\eta_k(t))\ket{\phi_k(t)} , \quad\quad
i\frac{\partial}{\partial t} \ket{\phi_{\bar k}(t)} =
(h(t)-\eta_{\bar k}(t))\ket{\phi_{\bar k}(t)} , \\
\label{drho_dt}
&&
i\frac{d}{dt}\rho_k(t) =
\kappa_k(t) \Delta_k^*(t)
-\kappa_k^*(t) \Delta_k(t) , \\
\label{dkappa_dt}
&&
i\frac{d}{dt}\kappa_k(t) =
\left(
\eta_k(t)+\eta_{\bar k}(t)
\right) \kappa_k(t) +
\Delta_k(t) \left( 2\rho_k(t) -1 \right) .
\end{eqnarray}
\end{subequations}
These basic equations determine the time evolution of
the canonical states, $\ket{\phi_k(t)}$ and $\ket{\phi_{\bar k}(t)}$,
their occupation, $\rho_k(t)$, and pair probabilities, $\kappa_k(t)$.
The real functions of time, $\eta_k(t)$ and $\eta_{\bar k}(t)$,
are arbitrary and associated with the gauge degrees of freedom.
The time-dependent pairing gaps,
$\Delta_k(t)$, are defined 
in the same manner as the BCS pairing gap \cite{RS80} except for
the fact that the canonical pair of states are no longer
related to each other by time reversal.
It should be noted that the quantities in the Cb-TDHFB equations,
$(\rho, \kappa, \Delta)$, are not matrixes but are only their diagonal elements
with a single index for the canonical states $k$.

The Cb-TDHFB equations, \eqref{Cb-TDHFB},
are invariant with respect to the gauge transformation with
arbitrary real functions, $\theta_k(t)$ and $\theta_{\bar k}(t)$.
\begin{eqnarray}
\label{gauge_transf_1}
\ket{\phi_k}\rightarrow e^{i\theta_k(t)}\ket{\phi_k}
\quad &\mbox{and}& \quad
\ket{\phi_{\bar k}}\rightarrow e^{i\theta_{\bar k}(t)}\ket{\phi_{\bar k}}
\\
\label{gauge_transf_2}
\kappa_k\rightarrow e^{-i(\theta_k(t)+\theta_{\bar k}(t))}\kappa_k
\quad &\mbox{and}& \quad
\Delta_k\rightarrow e^{-i(\theta_k(t)+\theta_{\bar k}(t))}\Delta_k
\end{eqnarray}
simultaneously with
$$
\eta_k(t)\rightarrow \eta_k(t)+\frac{d\theta_k}{dt}
\quad \mbox{and} \quad
\eta_{\bar k}(t)\rightarrow \eta_{\bar k}(t)+\frac{d\theta_{\bar k}}{dt} .
$$
It is now clear that the arbitrary real functions,
$\eta_k(t)$ and $\eta_{\bar k}(t)$,
control time evolution of the phases of
$\ket{\phi_k(t)}$, $\ket{\phi_{\bar k}(t)}$, $\kappa_k(t)$, and
$\Delta_k(t)$.

\begin{wrapfigure}{r}{8cm}
\includegraphics[height=8cm,angle=-90]{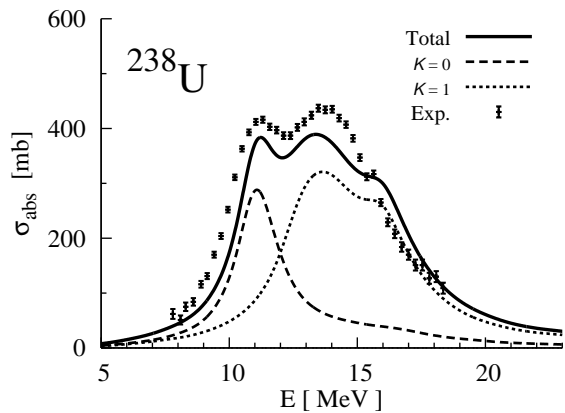}
\caption{Calculated (solid line) and experimental (symbols)
photoabsorption \cite{BF75} cross sections for $^{238}$U.
The dashed and dotted lines indicate the contribution from the dipole
oscillation parallel and perpendicular to the symmetry axis,
respectively.
}
\label{fig:238U}
\end{wrapfigure}
In addition to the gauge invariance,
the Cb-TDHFB equations possess the following properties.
\begin{enumerate}
\item Conservation law
 \begin{enumerate}
 \item Conservation of orthonormal property of the canonical states
 \item Conservation of average particle number
 \item Conservation of average total energy
 \end{enumerate}
\item The stationary solution corresponds to the HF+BCS solution.
\item Small-amplitude limit
 \begin{enumerate}
 \item The Nambu-Goldstone modes are zero-energy normal-mode solutions.
 \item If the ground state is in the normal phase, the equations are
         identical to the particle-hole, particle-particle, and hole-hole
         RPA with the BCS approximation.
 \end{enumerate}
\end{enumerate}

For numerical calculations, we extended the computational program of the
TDHF in the three-dimensional (3D) coordinate-space representation \cite{NY05}
to include the BCS-type pairing correlations.
The ground state is first constructed by the HF+BCS calculation.
Then, we add a weak impulse electric dipole field $V_{E1}(t)$ to
the ground state.
$V_{E1}(t)$ is chosen as
\begin{equation}
\label{V_E1}
V_{E1}(t) = -\eta F^{E1}_i \delta(t), \hspace{5mm}
F^{E1}_i=\sum_{i: {\rm protons}} (Ne/A) r_i 
       -\sum_{i: {\rm neutrons}} (Ze/A) r_i ,
\end{equation}
where $i=(x,y,z)$.
We solve the Cb-TDHFB equations in real time and real space.
To obtain the $E1$ strength function,
we calculate the time evolution of the expectation value
$\bra{\Psi(t)} F^{E1}\ket{\Psi(t)}$
under the external field $V_{E1}(t)$ with small $\eta$.
Then, the strength function $S(E1;E)$ can be obtained by the Fourier
transform of $\bra{\Psi(t)} F^{E1}\ket{\Psi(t)}$ with
an artificial damping factor with $\Gamma=1$ MeV:
$\bra{\Psi(t)} F^{E1}\ket{\Psi(t)} \rightarrow \bra{\Psi(t)} F^{E1}\ket{\Psi(t)}e^{-\Gamma t/2}$.

The calculated $E1$ strength distribution is transformed into
the photoabsorption cross section and shown in Fig. \ref{fig:238U},
for the $^{238}$U nucleus.
The ground state of $^{238}$U is deformed in the prolate shape with
the axial symmetry.
Thus, the photoabsorption peak is split into two peaks:
the lower peak represents an oscillation along the symmetry axis ($K=0$)
and the higher one corresponds to that perpendicular
the the symmetry axis ($K=1$).
The total cross section is the sum of these, which well agree with
the experimental data.
This can be another indication of the deformation of the $^{238}$U nucleus.

The computational task for solving Eq. (\ref{Cb-TDHFB}) is significantly smaller
than that for solving Eq. (\ref{TDHFB}).
This is due to the fact that the number of the calculated canonical
states $k,\bar k$ is roughly the same order as the particle number
($N\sim 10^2$),
which is much smaller than the dimension of the model space
($M\sim 10^4 - 10^5$).

\section{Summary}

The real-time calculation is a useful tool to calculate many-body
dynamics, especially when we are interested in bulk properties in
a wide range of energy, and when the complicated continuum boundary
condition is necessary to construct the energy eigenstates.
We have demonstrated its capability and usefulness in several examples,
including the fusion reaction and the response in nuclei.
The methodology is quite universal and applicable to a variety of
sub-fields in many-body quantum physics.

\ack

Most of the results presented in this paper has been achieved
by collaboration with
S. Ebata, M. Ito, K. Yabana, and K. Yoshida.
This work is supported by Grant-in-Aid for Scientific Research
in Japan (Nos. 21340073 and 20105003).

\section*{References}
\bibliographystyle{iopart-num}
\bibliography{nuclear_physics,chemical_physics,myself,current}

\end{document}